\documentclass{elsart}
\usepackage{graphicx,amssymb}
\usepackage[centertags]{amsmath}
\usepackage{amssymb,amsfonts}
\usepackage{latexsym}
\begin{document}
\begin{frontmatter}

\title{Some new static charged spheres}
\author{S.D. Maharaj\corauthref{cor}},
\corauth[cor]{Corresponding author.}
\ead{Maharaj@ukzn.ac.za}
\author{S. Thirukkanesh\thanksref{now}}
\thanks[now]{Permanent address: Department of
Mathematics, Eastern University, Chenkalady, Sri Lanka.}
\ead{thirukkanesh@yahoo.co.uk}
\address{Astrophysics and Cosmology Research Unit,
School of Mathematical Sciences, University of KwaZulu-Natal,
Private Bag X54001, Durban 4000, South Africa.}

\begin{abstract}
We present new exact solutions for the Einstein-Maxwell system in
static spherically symmetric interior spacetimes. For a particular
form of the gravitational potentials and the electric field
intensity, it is possible to integrate the system in closed form.
For specific parameter values it is possible to find new exact
models for the Einstein-Maxwell system in terms of elementary
functions. Our model includes a particular charged solution found
previously; this suggests that our generalised solution could be
used to describe a relativistic compact sphere. A physical analysis
indicates that the solutions describe realistic matter
distributions.
\end{abstract}
\end{frontmatter}

\section{Introduction}
Exact solutions of the Einstein-Maxwell system of field equations,
for spherically symmetric gravitational fields in static manifolds,
are necessary to describe charged compact spheres in relativistic
astrophysics. The solutions to the field equations generated have a
number of different applications in relativistic stellar systems. It
is for this reason that a number of investigations have been
undertaken on the Einstein-Maxwell equations in recent times. A
comprehensive review of exact solutions and criteria for physical
admissability is provided by Ivanov \cite{1}.  A general treatment
of nonstatic spherically symmetric solutions to the Einstein-Maxwell
system, in the case of vanishing shear was, performed by Wafo Soh
and Mahomed \cite{2} using symmetry methods. The uncharged case was
considered by Wafo Soh and Mahomed \cite{3} who show that all
existing solutions arise because of the existence of a Noether point
symmetry; the physical relevance of the solutions was investigated
by Feroze \emph{et al} \cite{4}. The matching of nonstatic charged
perfect fluid spheres to the Reissner-Nordstrom exterior metric was
pursued by Mahomed \emph{et al} \cite{5} who highlighted the role of
the Bianchi identities in restricting the number of solutions.

In this paper, we seek  a new class of solutions to the
Einstein-Maxwell system that satisfies the physical criteria. We
attempt to perform a systematic series analysis to the coupled
Einstein-Maxwell equations by choosing a rational form for one of
the gravitational potentials and a particular form for the charged
matter  distribution.  This approach produces a number of
difference equations, which we demonstrate can be solved
explicitly from first principles. A similar approach was used by
Thirukkanesh and Maharaj \cite{13}, Maharaj and Thirukkanesh
\cite{14}  and Komathiraj and Maharaj \cite{15}. They obtained
exact solutions by reducing the condition of pressure isotropy to
a recurrence relation with real and rational coefficients which
could be solved by mathematical induction. In this way new
mathematical and physical insights in the Einstein-Maxwell field
equations were generated.  An advantage of this approach is that
we generate new solutions to the Einstein-Maxwell system which
contain uncharged solutions found previously: Maharaj and Leach
\cite{16}, Tikekar \cite{17}, Durgapal and Bannerji \cite{18} and
John and Maharaj \cite{19}, amongst others.

We first express the Einstein-Maxwell system of equations for
static spherically symmetric line element as an equivalent system
using the Durgapal and Bannerji \cite{18} transformation in
Section 2. In Section 3, we choose specific forms for one of the
gravitational potentials and the electric field intensity, which
reduce the condition of pressure isotropy to a linear second order
equation in the remaining gravitational potential. We integrate
this generalised condition of isotropy equation using the method
of Frobenius  in Section 4. In general the solution will be given
in terms of special functions. We demonstrate that it is possible
to find two category of solutions in terms of elementary functions
by placing certain  restriction on the parameters. We regain known
charged Einstein-Maxwell models from our general class of models
in Section 5. In Section 6, we discuss the physical features of
the solutions found, plot the matter variables, and show that our
models are physically reasonable.

\section{The field equations}
We assume that the spacetime is spherically symmetric and static
which is consistent with the study of charged compact objects in
relativistic astrophysics. In Schwarzschild coordinates
$(t,r,\theta,\phi)$ the generic form of the line element is given
by
\begin{equation}
\label{eq:d1} ds^{2} = -e^{2\nu(r)} dt^{2} + e^{2\lambda(r)}
dr^{2} +
 r^{2}(d\theta^{2} + \sin^{2}{\theta} d\phi^{2}).
\end{equation}
The Einstein field equations for the line element (\ref{eq:d1}) can
be written as
\begin{subequations}
\begin{eqnarray}
\label{eq:d2a}
\frac{1}{r^{2}} \left[ r(1-e^{-2\lambda}) \right]' & = & \rho +\frac{1}{2}E^2, \\
- \frac{1}{r^{2}} \left( 1-e^{-2\lambda} \right) +
\frac{2\nu'}{r}e^{-2\lambda} & = & p -\frac{1}{2}E^2,\\
\label{eq:d2c} e^{-2\lambda} \left( \nu'' + \nu'^{2} +
\frac{\nu'}{r} -\nu'\lambda' - \frac{\lambda'}{r} \right) & = &
p+\frac{1}{2}E^2, \\
\label{eq:d2d}\sigma & =& \frac{1}{r^2}e^{\lambda}\left(r^2
E\right)',
\end{eqnarray}
\end{subequations}
for charged perfect fluids. The energy density $\rho$ and the
pressure $p$ are measured relative to the comoving fluid
4-velocity $u^a = e^{-\nu} \delta^a_0$ and primes denote
differentiation with respect to the radial coordinate $r$. The
quantities associated with the electric field are $E$, the
electric field intensity, and $\sigma$, the proper charge density.
In the system (\ref{eq:d2a})-(\ref{eq:d2d}), we are using units
where the coupling constant $\frac{8\pi G}{c^4}=1$ and the speed
of light $c=1$. This system of equations determines the behaviour
of the gravitational field for a charged perfect fluid source. A
different, but equivalent form of the field equations, can be
found if we introduce the transformation
\begin{equation}
\label{eq:d3} x = Cr^2,~~~~ Z(x)  = e^{-2\lambda(r)}, ~~~~
A^{2}y^{2}(x) = e^{2\nu(r)},
\end{equation}
where the parameters  $A$ and $C$ are arbitrary constants. Under
the transformation (\ref{eq:d3}) the system
(\ref{eq:d2a})-(\ref{eq:d2d}) has the equivalent form
\begin{subequations}
\begin{eqnarray}
\label{eq:d4a} \frac{1-Z}{x} - 2\dot{Z} & = & \frac{\rho}{C} +
 \frac{E^{2}}{2C}, \\
4Z\frac{\dot{y}}{y} + \frac{Z-1}{x} & = & \frac{p}{C} -
 \frac{E^{2}}{2C}, \\
\label{eq:d4c}4Zx^{2}\ddot{y} + 2 \dot{Z}x^{2} \dot{y} +
\left(\dot{Z}x -
Z + 1 - \frac{E^{2}x}{C}\right)y & = & 0, \\
\label{eq:d4d} \frac{\sigma^{2}}{C} & = & \frac{4Z}{x} \left(x
\dot{E} + E \right)^{2},
\end{eqnarray}
\end{subequations}
where dots denote differentiation with respect to $x$. This system
of equations governs the behaviour of the gravitational field for
a charged perfect fluid source. When $E=0$ the Einstein-Maxwell
equations (\ref{eq:d4a})-(\ref{eq:d4d}) reduce to the uncharged
Einstein equations for a neutral fluid. Equation (\ref{eq:d4c}) is
called the generalised condition of pressure isotropy and is the
fundamental equation that needs to be integrated to demonstrate an
exact solution to the Einstein-Maxwell system of equations
(\ref{eq:d4a})-(\ref{eq:d4d}).

\section{Choosing potentials}
Our objective is to find a new  class of solutions to  the
Einstein-Maxwell system by making explicit choices for the
gravitational potential $Z$ and the electric field intensity  $E$.
We make the choice for $Z$ as
\begin{equation}
\label{eq:d5}Z(x)=\frac{(1+a x)^{2}}{1+bx},
\end{equation}
where $a$ and $b$ are  real constants. Note that the choice
(\ref{eq:d5}) ensures that the gravitational potential
$e^{2\lambda}$ is regular and well behaved in the stellar interior
for a wide range of values of the parameters $a$ and $b$. In
addition, when $x=0$ then $Z=1$ which ensures that there is no
singularity at the stellar centre. A special case of (\ref{eq:d5})
was studied by Komathiraj and Maharaj \cite{20}. The choice
(\ref{eq:d5})  does produces charged and uncharged solutions which
are necessary for constructing  realistic stellar models. On
substituting (\ref{eq:d5}) in (\ref{eq:d4c}) we obtain
\begin{eqnarray}
 &&
4(1+ax)^{2}(1+bx)\ddot{y}+2(1+ax)[b(1+ax)-2(b-a)]\dot{y}\nonumber
\\
\label{eq:d6}&&
+\left[(a-b)^{2}-\frac{E^{2}(1+bx)^{2}}{Cx}\right]y=0,
\end{eqnarray}
which is a second order differential equation.

The differential equation (\ref{eq:d6}) may be solved if a
particular choice of the electric field intensity $E$ is made. For
our purpose we set
\begin{equation}
\label{eq:d7}\frac{E^{2}}{C}=\frac{\alpha a(b-a)x}{(1+bx)^{2}},
\end{equation}
where $\alpha$ is a constant. The electric field intensity
specified in (\ref{eq:d7}) vanishes at the centre of the sphere;
it is continuous and bounded in the stellar interior for wide
range of values of $x$. The quantity $E^2$ has positive values in
the interior of star for relevant choices of  the constants
$\alpha, a$ and $b$.  Therefore the form given in (\ref{eq:d7}) is
physically reasonable to study the behaviour of charged spheres.
With  the choice (\ref{eq:d7}) we can express (\ref{eq:d6}) in the
form
\begin{eqnarray}
 &&
4(1+ax)^{2}\left[b(1+a x)- (b-a)\right]\ddot{y} + 2 a (1+a
x)\left[b(1+a x)- 2(b-a)\right]\dot{y}\nonumber
\\
\label{eq:d8}&& + a(b-a)(b-a-\alpha a)y=0.
\end{eqnarray}
In  (\ref{eq:d8})  we assume that $a\neq 0$ and $a\neq b$ so that
the electric field intensity is present. When $\alpha=0$  there is
no charge.

\section{Solutions}
To find the solution of the Einstein-Maxwell system we need to
integrate the master equation (\ref{eq:d8}). We consider two cases
on the integration process: $\alpha=\frac{b}{a}-1$ and $\alpha
\neq \frac{b}{a}-1$.

\subsection{The case $\alpha=\frac{b}{a}-1$}
In this case  the differential equation (\ref{eq:d8}) becomes
\begin{equation}
\label{eq:d9}2(1+ax)\left[b(1+a x)- (b-a)\right]\ddot{y} +  a
\left[b(1+a x)- 2(b-a)\right]\dot{y}=0.
\end{equation}
Equation (\ref{eq:d9}) is easily integrable and the solution can
be written as
\begin{equation}
\label{eq:d10}y(x)=c_{1}\left(\sqrt{\frac{a(1+bx)}{b-a}}-\arctan\sqrt{\frac{a(1+bx)}{b-a}}\right)+c_{2},
\end{equation}
where $c_1$ and $c_2$ are constants of integration. Therefore, the
solution of the Einstein-Maxwell system
(\ref{eq:d4a})-(\ref{eq:d4d}) becomes
\begin{subequations}
\label{eq:d11}
\begin{eqnarray}
e^{2\lambda}&=&\frac{1+bx}{(1+ax)^{2}},\\
e^{2\nu}&=&A^{2}\left[c_{1}\left(\sqrt{\frac{a(1+bx)}{b-a}}
-\arctan\sqrt{\frac{a(1+bx)}{b-a}}\right)+c_{2}\right]^{2},\\
\frac{\rho}{C}&=& \frac{(b-2a)(6+bx)}{2(1+bx)^2}- \frac{a^2 x(11+6bx)}{2(1+bx)^2},\\
\frac{p}{C}&=&\frac{(2a-b)(2+bx)}{2(1+bx)^2}+\frac{a^2x(3+2bx)}{2(1+bx)^2}\nonumber\\
&& +\frac{2a
c_1(1+ax)\sqrt{\frac{a(1+bx)}{b-a}}}{c_1(1+bx)\left(\sqrt{\frac{a(1+bx)}{b-a}}
-\arctan\sqrt{\frac{a(1+bx)}{b-a}}\right)+c_{2}}, \\
\frac{E^{2}}{C}&=& \frac{(b-a)^{2}x}{(1+bx)^{2}}.
\end{eqnarray}
\end{subequations}
Observe that  because of the restrictions $\alpha=\frac{b}{a}-1$ and
$b\neq a$ the charged solution (\ref{eq:d11}) does not have an
uncharged limit. Therefore this solution models a sphere that is
always charged and cannot attain a neutral state.  Note that the
solution (\ref{eq:d11}) is expressed in a simple form in terms of
elementary functions which facilitates a physical analysis of the
matter and gravitational variables.

\subsection{The case $\alpha \neq \frac{b}{a}-1$}
With $\alpha \neq \frac{b}{a}-1$, equation (\ref{eq:d8}) is
difficult to solve. Consequently we introduce the transformation
\begin{equation}
\label{eq:d12} y=(1+ ax)^{d}U(1+a x),
\end{equation}
where $U$ is a function of $(1 + ax)$ and $d$ is constant. With
the help of (\ref{eq:d12}), the differential equation
(\ref{eq:d8}) can be written as
\begin{eqnarray}
 &&4 (1+ ax)^2 \left[b(1+a x)-
(b-a)\right]\ddot{U}\nonumber\\
&&+ 2 (1+a
x)\left[b(4d+1)(1+ax)-2(2d+1)(b-a)\right]\dot{U}\nonumber \\
\label{eq:d13}&&+\left[2bd
(2d-1)(1+ax)-(b-a)\left(\frac{b}{a}-1-\alpha-4d^2\right)\right]U=0.
\end{eqnarray}
Note that there is substantial simplification if we take
\[\frac{b}{a}-1-\alpha=4d^{2}.\]
Then (\ref{eq:d13}) becomes
\begin{eqnarray}
\label{eq:d14}&&2(1+ax)\left[(1+ax)-\frac{(b-a)}{b}\right]\ddot{U}\nonumber\\
&&+\left[(4d+1)(1+ax)-2(2d+1)\frac{(b-a)}{b}\right]\dot{U}+d(2d-1)U=0,
\end{eqnarray}
where $b \neq 0$. We observe that the point $1+ax=\frac{b-a}{b}$
is a regular singular point of the differential equation
(\ref{eq:d14}). Therefore,  the solution of the differential
equation (\ref{eq:d14}) can be written in the form of an infinite
series by the method of Frobenius:
\begin{equation}
\label{eq:d15}U=\sum_{i=0}^{\infty}c_{i}\left[(1+ax)-\frac{(b-a)}{b}\right]^{i+r},~
c_{0}\neq0,
\end{equation}
where $c_{i}$ are the coefficients of the series and $r$ is the
constant. To complete the solution we need to find the
coefficients $c_{i}$ as well as the parameter $r$ explicitly.

The indicial equation determines the value of $r$ from
\[c_{0}r(2r-3)=0. \]
As $c_{0}\neq 0$ we must have $r=0$ or $r=3/2$. We can express the
 structure for the general coefficient $c_i$ in terms of the
leading coefficient $c_{0}$ as
\begin{equation}
\label{eq:d16}c_{i}=\left(\frac{b}{a-b}\right)^{i}\prod_{p=1}^{i}\frac{[(p+r-1)(2p+2r+4d-3)
+d(2d-1)]}{(p+r)(2p+2r-3)}c_{0},
\end{equation}
where  the conventional symbol $\prod$  denotes multiplication. We
can verify the result (\ref{eq:d16}) using mathematical induction.
We can now generate two linearly independent solutions to
(\ref{eq:d14}) with the help of (\ref{eq:d15}) and (\ref{eq:d16}).
For the parameter value $r=0$, we obtain the first solution
\begin{eqnarray}
 U_{1}&=&c_{0}\left[1+ \sum_{i=1}^{\infty}\left(
\frac{b}{a-b}\right)^{i}\prod_{p=1}^{i}
\frac{[(p-1)(2p+4d-3)+d(2d-1)]}{p(2p-3)} \right.\times
\nonumber\\
&&\left. \left[(1+ax)-\frac{(b-a)}{b}\right]^{i}\right] \nonumber.
\end{eqnarray}
For the parameter value $r=3/2$, we obtain the second solution
\begin{eqnarray}
&&U_{2}=c_{0}\left[(1+ax)-\frac{(b-a)}{b}\right]^{3/2}
\left[1+ \right.\nonumber\\
&&\left.
\sum_{i=1}^{\infty}\left(\frac{b}{a-b}\right)^{i}\prod_{p=1}^{i}
\frac{[(2p+1)(p+2d)+d(2d-1)]}{p(2p+3)}\left[(1+ax)-\frac{(b-a)}{b}\right]^{i}\right].\nonumber
\end{eqnarray}
Since the functions $U_{1}$ and $U_{2}$ are  linearly independent
we have found the general solution to (\ref{eq:d14}). Therefore,
the solutions to the differential equation (\ref{eq:d8}) are
\begin{eqnarray}
&& y_{1}(x)=c_{0}(1+ax)^{d}\left[1+
 \sum_{i=1}^{\infty}\left(\frac{b}{a-b}\right)^{i} \times\right. \nonumber\\
\label{eq:d17}&& \left.\prod_{p=1}^{i}
\frac{[(p-1)(2p+4d-3)+d(2d-1)]}{p(2p-3)}\left[(1+ax)-\frac{(b-a)}{b}\right]^{i}\right]
\end{eqnarray}
and
\begin{eqnarray}
&&y_{2}(x)=c_{0}(1+ax)^{d}
\left[(1+ax)-\frac{(b-a)}{b}\right]^{3/2}\left[1+
\sum_{i=1}^{\infty}\left(\frac{b}{a-b}\right)^{i}\times \right. \nonumber\\
\label{eq:d18}&&\left. \prod_{p=1}^{i}
\frac{[(2p+1)(p+2d)+d(2d-1)]}{p(2p+3)}\left[(1+ax)-\frac{(b-a)}{b}\right]^{i}\right].
\end{eqnarray}
Thus the general solution to the differential equation
(\ref{eq:d6}), for the choice of the electric field (\ref{eq:d7}),
is given by
\begin{equation}
\label{eq:d19}y(x)=A_{1}y_{1}(x)+A_{2}y_{2}(x),
\end{equation}
where $A_{1}$ and $A_{2}$ are arbitrary constants and
$d^{2}=\left(\frac{b}{a}-1-\alpha \right)/4$.
 From (\ref{eq:d19}) and (\ref{eq:d4a})-(\ref{eq:d4d}), the exact
 solution of the Einstein-Maxwell system becomes
\begin{subequations}
\label{eq:d20}
\begin{eqnarray}
\label{eq:d20a}
e^{2\lambda}&=&\frac{1+bx}{(1+ax)^{2}},\\
\label{eq:d20b}
e^{2\nu}&=&A^{2}y^{2},\\
\label{eq:d20c}\frac{\rho}{C}&=&\frac{(3+bx)(b-2a)}{(1+bx)^2}-
\frac{a^2 x (5+3bx)}{(1+bx)^2}-\frac{\alpha a(b-a)x}{2(1+bx)^2},\\
\label{eq:d20d}\frac{p}{C}&=&4 \frac{(1+ax)^2}{(1+bx)}
\frac{\dot{y}}{y}+\frac{a(2+ax)-b}{(1+bx)}+\frac{\alpha a(b-a)x}{2(1+bx)^2},\\
\frac{E^{2}}{C}&=& \frac{\alpha a(b-a)x}{(1+bx)^2}.
\end{eqnarray}
\end{subequations}
We believe that this is a new solution to the Einstein-Maxwell
system. In general the models in (\ref{eq:d20}) cannot be
expressed in terms of elementary functions as the series in
(\ref{eq:d17}) and (\ref{eq:d18}) do not terminate. Consequently
the solution will be given in terms of special functions.
Terminating series are possible for particular values of $a$ and
$b$ as we show in the next section.

\section{Elementary functions}
It is possible to generate exact solutions in terms of elementary
functions from the series in (\ref{eq:d19}). This is possible for
specific values of the parameters $a, b$ and $\alpha$ so that the
series  (\ref{eq:d17}) and (\ref{eq:d18}) terminate. Consequently
two categories of solutions are obtainable  in terms of elementary
functions by placing restrictions on the quantity $\frac{b}{a}-1
-\alpha$. We can express the first category of solution, in terms
of the variable $r$, as
\begin{eqnarray}
\label{eq:d21}
&&y_1(x)=A_{1}\frac{1}{(1+ax)^{n}}\times \nonumber\\
&&\sum_{i=0}^{n}(-1)^{i-1}\left(\frac{b}{b-a}\right)^i
\frac{(2i-1)}{(2i)!(2n-2i+1)!}\left[(1+ax)-\frac{(b-a)}{b}\right]^{i}\nonumber\\
&&+A_{2}\frac{1}{(1+ax)^{n}}\left[(1+ax)-\frac{(b-a)}{b}\right]^{3/2}\times \nonumber\\
&&\sum_{i=0}^{n-1}\left(\frac{b}{a-b}\right)^{i}\frac{(i+1)}{(2i+3)!(2n-2i-2)!}
\left[(1+ax)-\frac{(b-a)}{b}\right]^{i},
\end{eqnarray}
where $\frac{b}{a}-1-\alpha=4n^{2}$ relates the constants
$a,b,\alpha$ and $n$. The second category of solution, in terms of
the variable $r$, is given by
\begin{eqnarray}
\label{eq:d22}&&y_2(x)=A_{1}\frac{1}{(1+ax)^{n-1/2}}\times
\nonumber\\
&&\sum_{i=0}^{n}(-1)^{i-1}\left(\frac{b}{b-a}\right)
\frac{(2i-1)}{(2i)!(2n-2i)!}\left[(1+ax)-\frac{(b-a)}{b}\right]^{i}\nonumber\\
&&+A_{2}\frac{1}{(1+ax)^{n-1/2}}\left[(1+ax)-\frac{(b-a)}{b}\right]^{3/2}\times
\nonumber\\
&&\sum_{i=0}^{n-2}\left(\frac{b}{a-b}\right)^{i}\frac{(i+1)}{(2i+3)!(2n-2i-3)!}
\left[(1+ax)-\frac{(b-a)}{b}\right]^{i},
\end{eqnarray}
where $ \frac{b}{a}-1-\alpha=4n(n-1)+1$ relates the constants
$a,b,\alpha$ and $n$. Thus we have extracted two classes of
solutions (\ref{eq:d21}) and (\ref{eq:d22}) to the
Einstein-Maxwell system in terms of elementary functions from the
infinite series solution (\ref{eq:d19}). This class of solution
can be expressed as combinations of polynomials and algebraic
functions. The simple form of (\ref{eq:d21}) and (\ref{eq:d22})
helps in the study of the physical features of the model.

From our general classes of solutions (\ref{eq:d21}) and
(\ref{eq:d22}), it is possible to generate particular solutions
found for charged stars previously . If we take $b=1$ and
$K=\frac{1-a}{a}$ then  it is easy to verify that the equation
(\ref{eq:d21}) becomes
\begin{eqnarray}
\label{eq:d23}
&&y_1(x)=D_{1}\left[\frac{K}{K+1+x}\right]^{n}~\sum_{i=0}^{n}(-1)^{i-1}
\frac{(2i-1)}{(2i)!(2n-2i+1)!}\left[\frac{1+x}{K}\right]^{i}\nonumber\\
&&+D_{2}\left[\frac{K}{K+1+x)}\right]^{n}\left[\frac{1+x}{K}\right]^{3/2}
\times
\nonumber\\
&&
\sum_{i=0}^{n-1}(-1)^{i}\frac{(i+1)}{(2i+3)!(2n-2i-2)!}\left[\frac{1+x}{K}\right]^{i},
\end{eqnarray}
where $K-\alpha=4n^{2}$, $D_1= \frac{A_1}{(1-a)^n}$ and
$D_2=\frac{A_2}{(1-a)^{n-3/2}}$. Also, equation (\ref{eq:d22})
becomes
\begin{eqnarray}
\label{eq:d24}&&y_2(x)=D_{1}\left[\frac{K}{K+1+x}\right]^{n-1/2}~\sum_{i=0}^{n}(-1)^{i-1}
\frac{(2i-1)}{(2i)!(2n-2i)!}\left[\frac{1+x}{K}\right]^{i}\nonumber\\
&&+D_{2}\left[\frac{K}{K+1+x}\right]^{n-1/2}\left[\frac{1+x}{K}\right]^{3/2}\times
\nonumber\\
&&
\sum_{i=0}^{n-2}(-1)^{i}\frac{(i+1)}{(2i+3)!(2n-2i-3)!}\left[\frac{1+x}{K}\right]^{i},
\end{eqnarray}
where $ K-\alpha=4n(n-1)+1$, $D_1= \frac{A_1}{(1-a)^{n- 1/2}}$ and
$D_2=\frac{A_2}{(1-a)^{n-2}}$. Thus we have regained the
Komathiraj and Maharaj \cite{20} charged model; our solutions
allow for a wider range of models for charged relativistic
spheres. We illustrate this feature with an example involving a
specific value for $n$. For example, suppose that $n=1$ then
$b=(5+\alpha)a$ and we get
\begin{equation}
\label{eq:d25} y=\frac{a_1 (7+\alpha+3(5+\alpha)ax)+a_2
(1+(5+\alpha)ax)^{3/2}}{1+ax}
\end{equation}
from (\ref{eq:d21}) where $a_1$ and $a_2$ are new arbitrary
constants. From (\ref{eq:d25}) and (\ref{eq:d4a})-(\ref{eq:d4d})
the solution to the Einstein-Maxwell system becomes
\begin{subequations}
\label{eq:d26}
\begin{eqnarray}
e^{2\lambda}&=&\frac{1+(5+\alpha)ax}{(1+ax)^{2}},\\
e^{2\nu}&=&A^{2}\left[\frac{a_1 (7+\alpha+3(5+\alpha)ax)+a_2
(1+(5+\alpha)ax)^{\frac{3}{2}}}{1+ax}\right]^{2},\\
\frac{\rho}{C}&=&\frac{a(3+\alpha-ax)}{1+(5+\alpha)ax}
+\frac{2a(1+ax)\left[3+\alpha
-(5+\alpha)ax\right]}{\left[1+(5+\alpha)ax\right]^2}
\nonumber\\
&& -\frac{\alpha a^2
(4+\alpha)x}{2\left[1+(5+\alpha)ax\right]^2},\\
\frac{p}{C}&=&\frac{2a(1+ax)}{\left[1+(5+\alpha)ax\right]} \times \nonumber\\
&& \frac{\left[4a_1(4+\alpha)
+a_2(1+(5+\alpha)ax)^{\frac{1}{2}}\left(13+3\alpha+(5+\alpha)ax\right)\right]}
{\left[a_1 \left(7+\alpha+3(5+\alpha)ax\right)
+a_2\left(1+(5+\alpha)ax\right)^{\frac{3}{2}}\right]}\nonumber\\
&& -\frac{a(3+\alpha-ax)}{1+(5+\alpha)ax}+\frac{\alpha a^2
(4+\alpha)x}{2\left[1+(5+\alpha)ax\right]^2},\\
\frac{E^2}{C}&=&\frac{\alpha a^2
(4+\alpha)x}{\left[1+(5+\alpha)ax\right]^2}.
\end{eqnarray}
\end{subequations}
Note that the solution of the form (\ref{eq:d26}) cannot be
regained from Komathiraj and Maharaj \cite{20} charged models
except for the value of $a=\frac{1}{(5+\alpha)}$. This indicates
that our model is the generalisation of Komathiraj and Maharaj
charged models with more general behaviour in the gravitational
and electromagnetic fields.

\section{Physical analysis}
In this section, we briefly consider the physical features of the
models generated in this paper. For the pressure to vanish at the
boundary $r=R$ in the solution (\ref{eq:d20}) we require $p(R)=0$
which gives the condition
\begin{equation}
\label{eq:d27} 4(1+aCR^2)^2
\left[\frac{\dot{y}}{y}\right]_{r=R}+a(2+aCR^2)-b +\frac{\alpha
a(b-a)CR^2}{2(1+bCR^2)}=0,
\end{equation}
where $y$ is given by (\ref{eq:d17})-(\ref{eq:d19}). This will
constrain the values of $a, b$ and $\alpha$. The solution of the
Einstein-Maxwell system for $r>R$ is given by the Reissner-Nordstrom
metric as
\begin{equation}
\label{eq:d28} ds^{2} = - \left( 1 - \frac{2m}{r} +
\frac{q^{2}}{r^{2}} \right)dt^{2}+
 \left( 1 - \frac{2m}{r} + \frac{q^{2}}{r^{2}} \right)^{-1}dr^{2} +
r^{2} \left( d\theta^{2} + \sin^{2}{\theta} d\phi^{2} \right),
\end{equation}
where $m$ and $q$ are the total mass and the charge of the star.
To match the potentials in (\ref{eq:d20}) to (\ref{eq:d28})
generates the relationships between the constants $A_1, A_2, a, b$
and $R$ as follows
\begin{subequations}
\label{eq:d29}
\begin{eqnarray}
 \left( 1 - \frac{2m}{R} + \frac{q^{2}}{R^{2}}
\right)&=&A^2 [A_1y_1(R)+A_2y_2(R)]^2,\\
\left( 1 - \frac{2m}{R} + \frac{q^{2}}{R^{2}}
\right)^{-1}&=&\frac{1+bCR^2}{(1+aCR^2)^2}.
\end{eqnarray}
\end{subequations}
The matching conditions (\ref{eq:d27}) and (\ref{eq:d29}) place
restrictions on the metric coefficients; however there are
sufficient free parameters to satisfy the necessary conditions that
arise for the model under study. Since these conditions are
satisfied by the constants in the solution a relativistic star of
radius $R$ is realisable.

From (\ref{eq:d20a}) and (\ref{eq:d20b}) we  easily observe that the
gravitational potentials $e^{2\lambda}$ and $e^{2\nu}$ are
continuous and well behaved for wide range of the parameters $a$ and
$b$. From (\ref{eq:d20c}), the variable $x$ can be expressed solely
in terms of the energy density $\rho$  as
\begin{eqnarray}
&&x= \frac{1}{2b(3a^2C+b\rho)} \left[b^2C -2b\rho -5a^2C-2abC \pm
\sqrt{(a-b)C} \times
\right.\nonumber\\
 && \left.\sqrt{\left[-27a^2 bC +a^3 C(25+6b\alpha)-b^2
 (bC+8\rho)+ab(3bC+8\rho+2b\alpha\rho)\right]}\right] \nonumber
 \end{eqnarray}
Hence, from (\ref{eq:d20d}) the isotropic pressure $p$ can be
written as a function of energy density $\rho$ only. Therefore the
solutions generated in this paper satisfy the barotropic equation of
state $p=p(\rho)$. Many of the solutions found previously do not
satisfy this desirable feature. We illustrate the graphical
behaviour of matter variables in the stellar interior for the
particular solution (\ref{eq:d26}). We assume that $a_1=-4.897,
a_2=C=1$ and $a=\alpha=1/4$ for simplicity, and we consider the
interval $0\leq r \leq 1$. To generate the plots for $\rho, p, E^2,
dp/d\rho$ and $p$ vs $\rho$, we utilised the software package
Mathematica. The behaviour of the energy density is plotted in Fig.
1. It is positive and monotonically decreasing towards the boundary
of the stellar object. In Fig. 2, we have plotted the behaviour of
matter pressure $p$, which is regular, monotonically decreasing and
becomes zero at the vacuum boundary of the stellar object. In Fig.
3, we describe the behaviour of the electric field intensity. It is
well behaved and a continuous function. In Fig. 4, we have plotted
the speed of sound $dp/d\rho$. We observe that $0\leq dp/d\rho \leq
1$ throughout the interior of the stellar object. Therefore the
speed of the sound is less than the speed of the light and causality
is maintained. In Fig. 5, we have plotted the pressure $p$ verses
the density $\rho$ and we find that this approximates a linear
function. This behaviour is to be expected as the gradients of $p$
and $\rho$ have similar profiles in the stellar interior. Thus we
have demonstrated that the particular solution satisfies the
requirements for a physically reasonable stellar interior in the
context of general relativity.

\vspace{2cm}
\begin{figure}
\begin{center}
\includegraphics*[width=6cm]{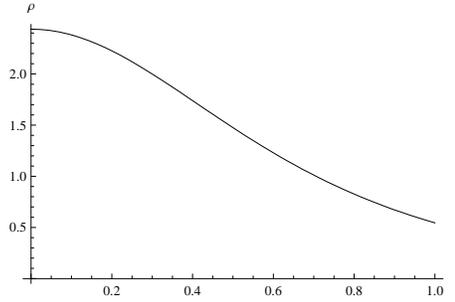}
\end{center}
\caption{Energy density.} \label{fig:1}
\end{figure}

\vspace{2cm}
\begin{figure}
\begin{center}
\includegraphics*[width=6cm]{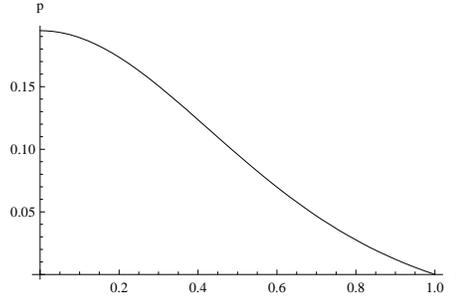}
\end{center}
\caption{Matter pressure.} \label{fig:2}
\end{figure}

\vspace{2cm}
\begin{figure}
\begin{center}
\includegraphics*[width=6cm]{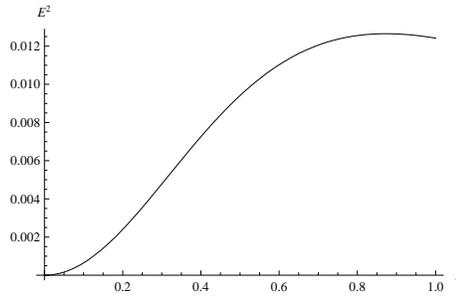}
\end{center}
\caption{Electric field intensity.} \label{fig:3}
\end{figure}

\vspace{2cm}
\begin{figure}
\begin{center}
\includegraphics*[width=6cm]{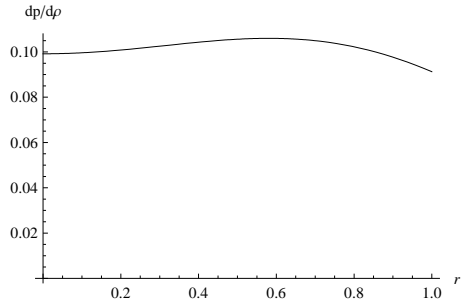}
\end{center}
\caption{Speed of sound $\frac{dp}{d\rho}.$} \label{fig:4}
\end{figure}

\vspace{2cm}
\begin{figure}
\begin{center}
\includegraphics*[width=6cm]{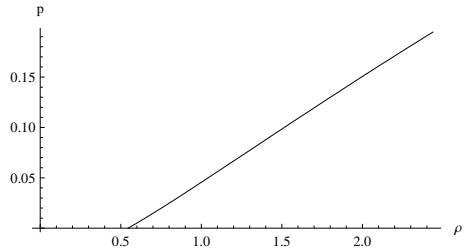}
\end{center}
\caption{Pressure vs Density} \label{fig:4}
\end{figure}

{\bf Acknowledgements}\\
ST thanks the National Research Foundation and the University of
KwaZulu-Natal for financial support, and is grateful to Eastern
University, Sri Lanka for study leave. SDM acknowledges that this
work is based upon research supported by the South African
Research Chair Initiative of the Department of Science and
Technology and the National Research Foundation.

\end{document}